\documentclass[12pt]{article}
\usepackage{graphics}
\usepackage{epsfig}
\usepackage{amsmath}
\title{Non-perturbative gluon evolution, squeezing, correlations and chaos in jets}
\author{V.I.~Kuvshinov, V.A.~Shaparau\\
\small\textit{Institute of Physics, National Academy
of Sciences of Belarus}\\
\small\textit{F. Skaryna av. 68, 220072 Minsk, Belarus}\\
\small \textit{E-mail: kuvshino@dragon.bas-net.by; shaporov@dragon.bas-net.by}\\
V.V.~Marmysh\\
\small\textit{Belorussian State University}\\
\small \textit{F. Skaryna av. 4, 220050 Minsk, Belarus}
}

\date{}

\begin{document}

\maketitle
\thispagestyle{empty}
\begin{abstract}
We study evolution of colour gluon states in isolated QCD jet at the non-perturbative stage. Fluctuations of gluons are less than those for coherent states under specific conditions. This fact suggests that there gluon squeezed states can arise. The angular and rapidity dependencies of the normalized second-order correlation
function for present gluon states are studied at this stage of jet evolution. It is shown that these new gluon states can have both sub-Poissonian and super-Poissonian statistics corresponding to, respectively, antibunching and bunching of gluons by analogy with squeezed photon states.

We investigate the possibility of coexisting both squeezing and chaos using Toda criterion and temporal correlator analysis. It is shown that these effects may coexist under some conditions.

\vspace{0.2cm}

\vspace{0.2cm}
\end{abstract}
\section{Introduction}

Many experiments at $ e^+e^-, p\bar p, ep $
colliders are devoted to hadronic jet physics, since detailed studies of
jets are important for better understanding and testing both
perturbative and non-perturbative QCD and also for finding
manifestations of new physics. Although the nature of jets is of a universal character, $ e^+e^- $- annihilation stands out among hard processes, since jet events admit a straightforward and clear-cut separation in this process. In the reaction $ e^+e^- \rightarrow hadron $ four evolution phases are recognized by various time and space scales.
These are (I) the production of a quark-antiquark pair: $e^{+}e^{-}\rightarrow
q\bar{q};$ (II) the emissions of gluons and quarks from primary partons --- perturbative evolution of the quark-gluon cascade; (III) the non-perturbative evolution and the hadronization of quarks and gluons; (IV) the decays of unstable particles.

The second phase of $ e^+e^- $- annihilation has been well understood and sufficiently accurate predictions for it have been obtained within the perturbative QCD (PQCD) \cite{PQCD-Coher,PQCD}. But predictions of the PQCD are limited by small effective coupling $\alpha(Q^2)<1$ and third phase is usually taken into account either through a constant factor which relates partonic features with hadronic ones (within local parton-hadron duality) or through the application of various phenomenological models of hadronization. As a consequence, theoretical predictions both for intrajet and for interjet characteristics remain unsatisfactory. For example, the width of the multiplicity distribution (MD) according to the predictions of PQCD is larger than the experimental one. The discrepancies between theoretical calculations and experimental data suggest that after perturbative stage the quark-gluon cascade undergoes non-perturbative evolution after that hadronization effects come into play. New gluon states, generated at the non-perturbative stage, contribute to various features of jets. For example, such a contribution to the multiplicity distribution can be in the form of the sub-Poissonian distribution \cite{Kuvsh,Kokoulina}. Therefore we must take into account both perturbative and non-perturbative stage of the jet evolution. 

Calculations performed within PQCD \cite{Malasa,Dremin} show that multiplicity distribution at the end of the perturbative cascade is close to a negative binomial distribution. At the same time, gluon MD in the range of the small transverse momenta (thin ring of jet) is Poissonian \cite{Soft}. Thus parton MD in the whole jet at the end of the perturbative cascade can be represented as a combination of Poissonian distributions  each of which corresponds to a coherent state.
Studying a further evolution of gluon states at the non-perturbative stage of jet evolution we obtain new gluon states that are squeezed states (SS) \cite{Nasa}-\cite{NPCS}. These states are formed as a result of non-perturbative self-interaction of the gluons expressed by nonlinearities of Hamiltonian. In this paper we prove that non-perturbative stage of jet evolution can be one of sources of a gluon SS by analogy with nonlinear medium for photon SS~\cite{QO_Japan}-\cite{Kilin}. Squeezed states posses uncommon properties: they display a specific behaviour of the factorial and cumulant moments \cite{PSS} and can have both sub-Poissonian and super-Poissonian statistics corresponding
to antibunching and bunching of photons. Moreover oscillatory behaviour MD of photon SS is differentiated from Poissonian and Negative binomial distributions (NBD). Because of analogy between photon and gluon, MD of gluon SS must have oscillations and using Local parton hadron duality (LPHD) we can compare derived gluon MD with hadron MD. It is clear that in this case behaviour of hadron MD in jet events is differentiated from NBD and this fact is confirmed by
experiments for $pp, p\overline{p}$-collisions~\cite{UA5}-\cite{OPAL}.

In series of works \cite{Aleks1}-\cite{Aleks3} it was shown that presence of a chaos amplifies effect of squeezing. At the same time condition of appearance of chaos is sufficiently understood,basically, in classical systems \cite{Licht}-\cite{Zasl}. Indeed, originally chaos was considered within classical and statistical physics. It was demonstrated that one of the causes of chaos is a local instability of a dynamical system \cite{Krylov} which can lead to mixing of trajectories in phase space and as result to non-regular behaviour of considered system \cite{Zasl}.

Large progress in understanding of chaos was achieved in semi-classical limit of
quantum mechanics \cite{Zasl_engl,Robnic}.
But not all criteria of the classical chaos are useful for quantum case.
In particular, by virtue of Heisenberg uncertainty relation
describing chaotic motion on the basis of exponentially
separation of neighboring trajectories is not possible in quantum
mechanics.  

A keen interest to chaos in field theories is connected with the facts that all four fundamental particle interactions have chaotic solutions \cite{Mand}.
 At the same time there are also some problems with definition of 
the chaos in the theory of field. In \cite{Kuzmin} it was shown that exponential decreasing of the Green function may serve as criterion of chaos both in the classical and quasiclassical theory of field.
Therefore we assume that conditions of appearance of chaos for quantum field systems, where squeezing exists, can find by studying features of the correlators of the field operators, which are analogue of the Green functions. 

Since the chaos phenomenon can be related with confinement \cite{Sav1} and with fractality for the factorial moments \cite{Kittel} then the question about condition of the appearance of the chaos in jet is important. In this connection we investigate SU(2)-jet model for the purpose of revealing of the local instability which can lead to chaos.

\section{The colour evolution of gluon states}
 The solution of the Schr$\mathrm{\ddot o}$dinger evolution equation for small time $t$
\begin{equation}\label{1H}
|f\rangle \simeq |i\rangle - \:i\, t\, H_g \,|i\rangle
\end{equation}
provides a possibility to observe an evolution of an initial state
vector $|i\rangle$ for small time.
Here the Hamiltonian\footnote{Here we suggest that a motion equation is satisfied, i.e. $\partial_\mu F^{\mu\nu}_a + g f_{abc}A_\mu^b F^{\mu\nu}_c = 0.$} $H_g$ has the form 
\begin{equation}\label{H_g}
  \hat H_g = H_0 + V,
\end{equation}
where $\displaystyle  H_0 =\frac12\int \bigl( \mathbf{E}_a \mathbf{E}_a +
\mathbf{B}_a \mathbf{B}_a\bigr)d^3\! x$ is the Hamiltonian of the "free"
gluons and the Hamiltonian of gluon self-interaction $V$ is
\begin{equation}
V = g\!\int \!\!\! f_{abc}
{\mathbf{E}}_{a}{\mathbf{A}}_{b}A^{0}_{c}\: d^{3}\! x -
\frac{g}{2}\!\int \!\!\! f_{abc}
{\mathbf{B}}_{a}[{\mathbf{A}}_{b}{\mathbf{A}}_{c}]d^{3}\! x
+\frac{g^{2}}{2}\!\int\!\!\!
\left(f_{abc}{\mathbf{A}}_{b}A^{0}_{c}\right)^{\! 2}\! d^{3}\! x +
\frac{g^{2}}{8}\!\int\!\!\!
\left(f_{abc}[{\mathbf{A}}_{b}{\mathbf{A}}_{c}]\right)^{2}\!
d^{3}\! x,
\end{equation}
${\mathbf{E}}_{a} = -{\mathbf{\nabla}}A^{0}_{a} -
\partial_{0}{\mathbf A}_{a}, {\mathbf{B}}_{a} =
[{\mathbf{\nabla}{A}} _{a}], A_{a}^\mu$ -- is the potential of gluon field ($\mu = \overline{0,4}, a = \overline{1,8}$). In the momentum representation the Hamiltonian of gluon self-interaction for the jet ring of thickness $d\theta$ can be represented in the form\footnote{By performing integration with respect to $\theta$, it is easy to obtain the gluon Hamiltonian for the whole jet; this Hamiltonian differs from the original one (\ref{v}) only by a factor. Hence, all of the further conclusions about the existence of squeezed states remain valid in this case.} \cite{acta, NPCS}
\begin{eqnarray}
\label{v} V&=&\displaystyle\frac{k_0^4}{4(2\pi)^3}\left
(1-\frac{q_0^2}{k_0^2}\right ) ^{3/2}\!\!\! g^2\, \pi\,
f_{abc}f_{ade}
  \Biggl\{ \left (2-\displaystyle\frac{q_0^2}{k^2_0}\right )
\left [a^{bcde}_{1212}+a^{bcde}_{1313}\right ]
+a^{bcde}_{2323}+\\
&&{}+\displaystyle\frac{\sin^2\theta}2\left (1-\displaystyle\frac
{q_0^2}{k^2_0} \right ) \!\!\left [2a^{bcde}_{2323}\,
-\,a^{bcde}_{1212}\, -\, a^{bcde}_{1313}\right ]\!
 \Biggr\}\sin\theta\:d\theta.\nonumber
\end{eqnarray}
Here $a^{bcde}_{lmlm}\!=\!a^{b+}_{l} a^{c+}_{m} a^{d}_{l}
a^{e}_{m}+ a^{b+}_l a^c_m a^{d+}_l a^e_m+a^b_l a^{c+}_m a^{d+}_l
a^e_m + h.c. ,\quad a^b_l(a^{b+}_l)$ is the operator annihilating (creating) a gluon of colour $b$ and vector component $l$, $q_0^2$ and $k_0$ are correspondingly the virtuality and energy of the gluon at the end of the perturbative cascade, g is the coupling constant, $f_{abc}$ stands for the structure constants of the $SU_{\!c}(3)$ group, $\theta $ is the angle between the jet axis and the momentum ${\mathbf k}$ ($0\le\theta\le\theta_{\max}, \theta_{\max} $ is half of the opening angle of the jet cone).

 Evolution of a single-gluon state vector $|a^b_l\rangle$
with colour $b$ and vector component $l$ has the evident form
\begin{equation}
|a^b_{\mathbf{l}}(t)\rangle \simeq |a^b_l(0)\rangle - \:i\,t\, C \,|a^b_l(0)\rangle,
\end{equation}
where quantity C depends from coupling constant, energy and
virtuality of gluon. As a result of single-gluon state evolution
the state vector with different colour is not appear. In this
respect evolution of gluon coherent state $|\alpha^b_l\rangle$ is more
interesting.
 For example, evolution of gluon coherent state with colour $b=1$ and vector
component $l=1$ in jet ring is the following
\begin{equation}
\begin{array}{l}
\mid\alpha^1_1 (t)\rangle\simeq\biggl\{ 1\, -\, 2it \pi\sin\theta\, d\theta
\Bigl(u_3\, +\, u_4\,\mid\alpha^1_1\mid^2\Bigr)\!\biggr\}|\alpha_1^1(0)\rangle-\,
2it\pi\sin\theta d\theta\, u_4\:\alpha^1_1  D
\bigl(\alpha^1_1\bigr) \mid a_1^1\rangle \\ \qquad{}\qquad\quad-\,2it
\pi\, u_2 \Bigl(1+u_1-\displaystyle\frac{u_1}{2}\sin^2\theta\Bigr)\sin\theta\, d\theta\,\left(\alpha^1_1\right)^2
\mathop{{\sum}'}\limits_{k=2}^7\sum\limits_{n=2}^{3}
\mid\alpha^1_1 (0),2a^k_n\rangle,
\end{array}
\end{equation}
where $\mid a^b_l(0)\rangle$ is a single gluon vector, $D(\alpha)=exp\{\alpha  a^+ - \alpha^* a\}$ is the unitary
displacement operator of amplitude $\alpha$,

\begin{equation}
\left.
\begin{array}{l}
\displaystyle \mathop{{\sum}'}\limits_{k=2}^{7}\bigl( \;\bigr) =
\mathop{\sum}\limits_{k=2}^{3}\bigl( \;\bigr)+ \frac14
\mathop{\sum}\limits_{k=4}^{7}\bigl( \;\bigr) ,\quad
 u_1
=\biggl(1-\displaystyle\frac{q_0^2}{k^2_0}\biggr),\quad
 u_2 =
\displaystyle\frac{k^4_0}{4(2\pi)^3}\frac{g^2}2 \sqrt{u_1^{3}},\\ \\
 u_3 =
\displaystyle\frac{15}2 k^3_0\,\sqrt{u_1}\,
\biggl(1+u_1+\frac{q_0^4}{2k^4_0}\biggr)+ \,24\,u_2
(3+2u_1),\\ \\
 u_4 = \displaystyle\frac{k^3_0}2 \sqrt{u_1}\biggl[\frac{q_0^4}{k^4_0}\,+\,
\biggl(1+u_1-\displaystyle\frac{q_0^4}{k^4_0}\biggr)\sin^2\theta\biggr]+\,6\,u_2[2(1+u_1)- u_1\sin^2\theta ].
\end{array}
\right\}
\end{equation}

 Analogously we can also investigate the evolution of gluon coherent
 states with any other colour charges and vector components.

 As a result the following conclusions have been obtained:\\
 1) for the initial vectors with the colour indices b=1,2,3 the vectors
with another colour indices k$=\overline{4,7}$ appear;\\
 2) if the initial vectors have the colour indices b=$\overline{4,7}$
then the new vectors with colour indices k=1,2,3,8 and the vectors
with the combination of the colour indices 3,8:
$\mid\alpha^{b}_{l},a^{3}_{l},a^{8}_{l}\rangle$ appear;\\
 3) as a result of the evolution of colour coherent state with b=8
the mixed colour states with colour indices 4,5,6,7 appear.

It is clear that namely the difference among the structure
constants of the $SU_c(3)$-group for different colour indices
leads to the different evolution of the corresponding colours.

Since product of the gluon coherent states with different colour and vector indices $\prod\limits_{c=1}^8\!\prod\limits_{l=1}^3\!\!\mid\!\alpha^c_l(0)\rangle$ corresponds to Poissonian distribution of the multimode gluon states in thin ring of jet, present state vector may be considered as initial state vector\footnote{For whole jet we have a superposition of these gluon coherent states with specific weights as initial state vector.} prepared by the perturbative stage. Therefore evolution problem of this initial state vector within a small interval of time $t$, which is defined according (\ref{1H}) as
\begin{equation}\label{f}
 |f\rangle=\prod\limits_{c=1}^8\!\prod\limits_{l=1}^3\!\!\mid\!
\alpha^c_l(t)\rangle\simeq
\prod\limits_{c=1}^8\!\prod\limits_{l=1}^3\!\!\mid\!
\alpha^c_l(0)\rangle - \:i t V
\,\prod\limits_{c=1}^8\!\prod\limits_{l=1}^3\!\!\mid\!
\alpha^c_l(0)\rangle,
\end{equation}
is actual. In this case the time is reckoned from the beginning of non-perturbative stage and the Hamiltonian of the gluon self-interaction $V$ in the jet ring is determined by formula (\ref{v}). The explicit form of the evolved state vector $|f\rangle$ is given in the Appendix A.

Because of producing of new evolved mixed state vectors the
question, connected with identification of these states, arises, in
particular, about fulfillment of squeezing condition
for these new states.

\section{Squeezed gluon states in QCD jet}
The Hamiltonian of the gluon self-interaction $V$ in the jet ring (\ref{v}) includes the squares of the creation and annihilation operators for gluons with specified colour and vector indices. As is known from quantum mechanics and quantum optics, the presence of such structure in the Hamiltonian and, consequently, in the evolution operator is a necessary condition for emergence of squeezed states \cite{QO_Cambr}, since the unitary squeezing operator involves quadratic combinations of the creation and annihilation operators:
\begin{equation}
\label{ss_operator}
 S(z)= \exp\Bigl\{\frac{z^{*}}{2} a^{2} - \frac{z}{2}
(a^{+})^{2}\Bigr\},
\end{equation}
where $z=r e^{i\delta}$ is an arbitrary complex number, $r$ is a squeeze factor, phase $\delta$ defines the direction of
squeezing maximum \cite{QO_Cambr}.

In order to verify whether the final gluon state vector describes the single-mode SS, it is necessary to introduce the phase-sensitive Hermitian operators $(X^b_l)_1=\big[a^b_l\, +\, (a^b_l)^+\big] /2 $ and
$(X^b_l)_2=\big[a^b_l\, -\, (a^b_l)^+\big] /2i\,$ by analogy with quantum optics and to establish conditions under which the variance of one of them can be less than the variance of a coherent state.

Mathematically, the condition of squeezing is expressed in the form of the inequalities \cite{Kilin} 
\begin{eqnarray*}
\Bigl\langle\left( {\Delta}(X^b_l)_{1\atop 2}\right)^2
\Bigr\rangle = \Bigl\langle N\left({\Delta}(X^b_l)_{1\atop
2}\right)^2 \Bigr\rangle + \frac14\, <\, \frac14
\end{eqnarray*}
or
\begin{eqnarray}
\label{sq_cond} \Bigl\langle N\left({\Delta}(X^b_l)_{1\atop
2}\right)^2 \Bigr\rangle <\, 0.
\end{eqnarray}
Here $N$ is the normal-ordering operator such as
\begin{eqnarray}
\label{form_N} \Bigl\langle N\left(\Delta(X^b_l)_{1\atop
2}\right)^2 \Bigr\rangle &=&\frac14
\Biggl\{\,\pm\,\left[\Bigl\langle\Bigl(a^b_l\Bigr)^2\Bigr\rangle
-\Bigl\langle a^b_l\Bigr\rangle^2\right]
\pm \left[\Bigl\langle\Bigl(a^{b+}_l\Bigr)^2\Bigr\rangle
-\Bigl\langle a^{b+}_l\Bigr\rangle^2\right]\\&&\qquad+
2\left[\Bigl\langle a^{b+}_l a^b_l\Bigr\rangle\, - \Bigl\langle
a^{b+}_l\Bigr\rangle \Bigl\langle a^b_l\Bigr\rangle\right]\!
\Biggr\}.\nonumber
\end{eqnarray}
The expectation values of the creation and annihilation operators for gluons with specified colour and vector components is taken for the vector $\prod\limits_{c=1}^8\!\prod\limits_{l=1}^3\!\!\mid\!
\alpha^c_l(t)\rangle$ (A.1). 

Let us consider the specific case where the colour index is $b=1$ and the vector index $l$ is arbitrary. Then we have
\begin{eqnarray}\label{sq_cond1}
\Bigl\langle N\left({\Delta}(X_l^1)_{1\atop 2}\right)^2
\Bigr\rangle\! =&\pm \: 4\pi\, u_2\, t\, \sin\theta\,d\theta
\biggl\{\!(1+u_1)\Bigl[
\delta_{l1}(Z_{33}+Z_{22})+(1-\delta_{l1})Z_{11}
\Bigr]+\Bigl[\delta_{l2} Z_{33}+
\delta_{l3}Z_{22}\Bigr]\nonumber\\&+u_1\sin^2\theta\:\Bigl[-\displaystyle\frac12\delta_{l1}
(Z_{22}+Z_{33})+\delta_{l2}(Z_{33}-\frac12
Z_{11})+\delta_{l3}(Z_{22}-\frac12 Z_{11} )\Bigr] \biggr\}.
\end{eqnarray}
Here $ Z_{mn}=\mathop{{\sum}'}\limits_{k=2}^7\bigl\langle
(X_m^k)_1\bigr\rangle \bigl\langle (X_n^k)_2\bigr\rangle\quad
(m,n=1,2,3).$
Since at small value of the squeeze factor we have
\begin{equation}\label{r}
  r \cos\delta=\mp\,2\,\Bigl\langle N\left({\Delta}(X)_{1\atop 2}\right)^2
\Bigr\rangle, 
\end{equation}
then taking into account formula (\ref{sq_cond1}) the expression (\ref{r}) can be rewrite in the form
\begin{equation}\label{r1}
\begin{array}{l}
r_l^1 \cos\delta = - \, 8\pi\, u_2\, t\, \sin\theta\,d\theta
\biggl\{\!(1+u_1)\Bigl[
\delta_{l1}(Z_{33}+Z_{22})+(1-\delta_{l1})Z_{11}
\Bigr]
+ \delta_{l2} Z_{33}+\delta_{l3}Z_{22}\\\qquad\qquad\quad + u_1\sin^2\theta\:\Bigl[-\displaystyle\frac12\delta_{l1}
(Z_{22}+Z_{33})+\delta_{l2}(Z_{33}-\frac12
Z_{11})+\delta_{l3}(Z_{22}-\frac12 Z_{11} )\Bigr] \biggr\}.
\end{array}
\end{equation}
Evidently that $\Bigl\langle N\left({\Delta}(X)_{1\atop 2}\right)^2
\Bigr\rangle\! \neq 0$ if $r\neq 0$ and $\delta\neq \displaystyle\frac\pi2, \frac{3\pi}2.$
In final state being consideration fluctuations of one  of the squared components of the gluon field, ${\Delta}(X^1_l)_{2}$, are less than those in the initial coherent state under the following conditions: $\bigl\langle (X^k_m)_1\bigr\rangle <0, \bigl\langle
(X^k_m)_2\bigr\rangle <0$ or
 $\bigl\langle (X^k_m)_1\bigr\rangle >0$ and $\bigl\langle
(X^k_m)_2\bigr\rangle >0$\quad $(k\neq 1, m\neq l)$ then, as following from (\ref{r1}), $\displaystyle\frac\pi2 < \delta < \frac{3\pi}2$. In this case we have phase-squeezed gluon states by analogy with quantum optics \cite{QO_Walls, QO_Cambr}. If the conditions $\bigl\langle (X^k_m)_1\bigr\rangle >0, \bigl\langle
(X^k_m)_2\bigr\rangle <0$ or
 $\bigl\langle (X^k_m)_1\bigr\rangle <0$ and $\bigl\langle
(X^k_m)_2\bigr\rangle >0$\quad $(k\neq 1, m\neq l, \displaystyle -\frac\pi2 < \delta < \frac{\pi}2)$  are satisfied, fluctuations in another squared component of the gluon field, ${\Delta}(X^1_l)_{1}$, will be less in the final state vector $\prod\limits_{c=1}^8\!\prod\limits_{l=1}^3\!\!\mid\!
\alpha^c_l(t)\rangle$ than in the coherent state. In this case we arrive at the amplitude-squeezed states (as in the case of photons \cite{QO_Walls, QO_Cambr}).

Rewriting the expression (\ref{r1}) in terms of the amplitude and phase of the gluon coherent states ($\alpha^b_l=\mid\!\alpha^b_l\!\!\mid e^{i\gamma_l^b}$)
\begin{eqnarray}\label{r2}
r_l^1 \cos\delta =& - \, 8\pi\, u_2\, t\, \sin\theta\,d\theta\mathop{{\sum}'}\limits_{k=2}^7
\biggl\{
\Bigl(1 + u_1 - \displaystyle\frac{u_1}2\sin^2 \theta\Bigr)\Bigl[\delta_{l1}\sum\limits_{n=2}^3 |\,\alpha_n^k|^2 \sin (2\gamma_n^k) + (1-\delta_{l1})|\,\alpha_1^k|^2 \sin (2\gamma_1^k)\Bigr]\nonumber\\&+\,(1+u_1 \sin^2 \theta)\Bigl[\delta_{l2}|\,\alpha_3^k|^2 \sin (2\gamma_3^k) + \delta_{l3}|\,\alpha_2^k|^2 \sin (2\gamma_2^k)\Bigr]
\biggr\},
\end{eqnarray}
we see that effect of the single-mode squeezing is absent ($r_l^1 \cos\delta = 0$) then the initial gluon coherent fields are either real ($\gamma_n^k=0, n\neq l, k\neq 1$) or imaginary ($\gamma_n^k=\pi/2, n\neq l, k\neq 1$).
Similar conclusions will also be valid for a gluon field featuring other colour indices.

Thus, the vector $\prod\limits_{c=1}^8\!\prod\limits_{l=1}^3\!\!\mid\!
\alpha^c_l(t)\rangle$ describes the squeezed state of gluons that are produced at the non-perturbative stage of the jet evolution within a small interval of time $t$. Here, the corresponding fluctuations of the squared components of the gluon field will be less than those in the case of the initial coherent state.

It should be noted that the Hamiltonian of the three-gluon self-interaction which in momentum representation has the next form
\begin{equation}\label{three}
\begin{array}{l}
V' = \displaystyle\frac{ig}{(2\pi)^{3/2}}\, f_{abc}\int\prod\limits_{i=1}^{3}\Bigl(\frac{d^3 k_i}{\sqrt{2 k_{0i}}}\Bigr)\biggl\{\Bigl[(\vec{a}_a\vec{a}_b)(\vec{k}_1\vec{a}_c) + [\frac{k_{01}}{k_{03}} \vec{a}_a - \frac{\vec{k}_{1}}{k_{03}k_{01}}(\vec{k}_1 \vec{a}_a)]\vec{a}_b(\vec{k}_3 \vec{a}_c)\Bigr]e^{-i(k_{01}+k_{02}+k_{03})t}\\ \times\delta(\vec{k}_1+\vec{k}_2+\vec{k}_3)+ 
\Bigl[(\vec{a}_a\vec{a}_b^+)(\vec{k}_1\vec{a}_c) + \displaystyle[\frac{k_{01}}{k_{03}} \vec{a}_a - \frac{\vec{k}_{1}}{k_{03}k_{01}}(\vec{k}_1 \vec{a}_a)]\vec{a}_b^+(\vec{k}_3 \vec{a}_c)\Bigr]e^{-i(k_{01}-k_{02}+k_{03})t} \delta(\vec{k}_1-
\vec{k}_2+\vec{k}_3)\\
+\Bigl[(\vec{a}_a\vec{a}_b)(\vec{k}_1\vec{a}_c^+) + \displaystyle[\frac{k_{01}}{k_{03}} \vec{a}_a - \frac{\vec{k}_{1}}{k_{03}k_{01}}(\vec{k}_1 \vec{a}_a)]\vec{a}_b(\vec{k}_3 \vec{a}_c^+)\Bigr]e^{-i(k_{01}+k_{02}-k_{03})t} \delta(\vec{k}_1+
\vec{k}_2-\vec{k}_3)\\
+\Bigl[(\vec{a}_a\vec{a}_b^+)(\vec{k}_1\vec{a}_c^+) + \displaystyle[\frac{k_{01}}{k_{03}} \vec{a}_a - \frac{\vec{k}_{1}}{k_{03}k_{01}}(\vec{k}_1 \vec{a}_a)]\vec{a}_b^+(\vec{k}_3 \vec{a}_c^+)\Bigr]e^{-i(k_{01}-k_{02}-k_{03})t} \delta(\vec{k}_1-
\vec{k}_2-\vec{k}_3) - h.c.
\biggr\}
\end{array}
\end{equation}
don't lead to squeezing effect.
In fact, rewriting the expression for $\Bigl\langle N\left({\Delta}(X_l^h)_{1\atop 2}\right)^2\Bigr\rangle$ (\ref{form_N}) for small time as
\begin{equation}\label{sq_cond2}
\begin{array}{l}
\Bigl\langle N\left({\scriptstyle\triangle}(X_{l}^h)_{1\atop 2}\right)^2
\Bigr\rangle = 
\mp\,\displaystyle \frac{i\,t}{4}\biggl\{\langle\alpha\,|\,[a^h_{l}\left(k\right),[a^h_{l}\left(k\right), V ]\,]|\alpha\rangle -  \langle\alpha\,|[\,[V, a^{+h}_{l}\left(k\right)], a^{+h}_{l}\left(k\right)]\alpha\rangle\biggr\},
\end{array}
\end{equation}
it can be shown that
$
[a^h_{l}\left(k\right),[a^h_{l}\left(k\right), V']\,]=0,\;
[\,[V', a^{+h}_{l}\left(k\right)], a^{+h}_{l}\left(k\right)]=0
$ since $f_{hhb}=0,$ that is the squeezing condition (\ref{sq_cond}) does not hold.

\section{Correlation functions for squeezed gluon states}

The behaviour of a correlation function can serve one of a criterion of the existence of squeezed gluon states. It is common to define the normalized second-order correlation function as \cite{Kittel}
\begin{eqnarray}
K_{(2)}(\theta_1,\theta_2)&=&\frac{C_{(2)}(\theta_1,\theta_2)}{\rho_1(\theta_1)
\rho_1(\theta_2)},
\end{eqnarray}
 $C_{(2)}(\theta_1,\theta_2)=
\rho_2(\theta_1,\theta_2)-\rho_1(\theta_1)\rho_1(\theta_2)$, with
$\rho_2(\theta_1,\theta_2)\,(\rho_1(\theta)\,)$ being the two-particle (single-particle) inclusive distribution. Then for gluons with a colour $b$ and a vector component $l$ we can write
\begin{eqnarray}
K^b_{l(2)}(\theta_1,\theta_2)&=&\frac{\rho^b_{l(2)}(\theta_1,\theta_2)}
{\rho^b_{l(1)}(\theta_1)\rho^b_{l(1)}(\theta_2)} - 1.
\end{eqnarray}
At the same time, we have
\begin{equation}
\label{cor_func}
 \int\limits_{\Omega}\!\rho_1(\theta)\,d\theta =\;
\langle n\rangle\;=\;\langle a^+ a\rangle\; =\int\limits_{\Omega}\!\!\langle f(\theta,t)|
a^+ a|f(\theta,t)\rangle d\theta,
\end{equation}
where $|f(\theta,t)\rangle$ is the final state vector. From (\ref{cor_func}) we find that the single- and two-particle inclusive distributions can be represented as
\begin{equation}\label{ro}
\left.
\begin{array}{l}
\rho_1(\theta)= \langle f(\theta,t)|a^+ a|f(\theta,t)\rangle,\\ \\
\rho_2(\theta_1,\theta_2)=\langle f(\theta_2,t),f(\theta_1,t)| a^+
a^+ a\, a |f(\theta_1,t),f(\theta_2,t)\rangle.
\end{array}
\!\!\!\right\}
\end{equation}
By taking the expectation values over the vector\footnote{That this vector also describes squeezed gluon states can be proven by verifying the squeezing condition (\ref{sq_cond}).} $\prod\limits_{c=1}^8\prod\limits_{l=1}^3\!\!\mid\!\alpha
^c_l(\theta_1,t), \alpha ^c_l(\theta_2,t)\rangle$, we obtain the explicit form of the normalized second-order correlation function for squeezed gluon states:
\begin{equation}
\label{corforgl}
 K^{b}_{l(2)}(\theta_1,\theta_2) = -\,M_1
(\theta_1,\theta_2)/\{\mid\alpha^b_l\mid^4-2\mid\alpha^b_l\mid^2M_1
(\theta_1,\theta_2)+ M_2 (\theta_1,\theta_2)\}.
\end{equation}
For the colour index $b=1$ and an arbitrary vector component $l$ we have here
\begin{eqnarray}
M_1(\theta_1,\theta_2) &=&
24\,t\,u_2\:\pi\mid\alpha\mid^2\mid\beta\mid^2
\sin\left(\delta+\displaystyle\frac{\pi}{2}\right)
\Bigl\{(1+\delta_{l1})(2+u_1-\delta_{l1})(\sin\theta_1\,+\sin\theta_2)\nonumber\\&&{}
 -
\displaystyle\frac12\,u_1\:(3\delta_{l1} - 1)
(\sin^3\theta_1\,+\sin^3\theta_2) \Bigr\},
\end{eqnarray}
\begin{eqnarray}
M_2(\theta_1,\theta_2) &=&
80\,t\,u_2\:\pi\mid\alpha\mid^3\mid\beta\mid^3
\sin\left(\displaystyle\frac{\delta}{2}+\displaystyle\frac{\pi}{4}\right)
\Bigl\{(1+\delta_{l1})(2+u_1-\delta_{l1})
 (\sin\theta_1\,+ \sin\theta_2)\nonumber\\&&{}-\displaystyle\frac12\,u_1\:(3\delta_{l1} - 1)
(\sin^3\theta_1\, +\sin^3\theta_2) \Bigr\}.
\end{eqnarray}
In deriving these formulae, for simplicity we assumed that $\alpha^1_l=\mid\!\!\alpha\!\!\mid
e^{i\gamma_1},$ $l=\mathrm{any}$, and $\alpha^b_l=\mid\!\!\beta\!\!\mid
e^{i\gamma_2}, $ when $b\ne 1$ and an arbitrary  $l$, $\gamma_1 - \gamma_2 = \delta/2 + \pi/4$.

Let us perform a comparative analysis of the correlation function (\ref{corforgl}) for gluon squeezed states and the corresponding function for photon squeezed states, which was thoroughly studied in quantum optics. 

In quantum optics, the normalized second-order correlation function is defined as \cite{Kilin}
\begin{eqnarray}
K_{l(2)} = g^{(2)}_{l} - 1 &=&\frac{\left\langle \hat a^+_l\,\hat
a^+_l\,\hat a_l\, \hat a_l\right\rangle}{\left\langle \hat
a^{1+}_l\,\hat a_l\right\rangle^2} - 1,
\end{eqnarray}
where the expectation values are taken over the evolved state vector at the instant $t$. If the correlation function is positive, occurs photon bunching (super-Poissonian distribution); otherwise ($K_{l(2)} < 0$), we have photon antibunching (sub-Poissonian distribution) \cite{QO_Walls, Kilin}. For a coherent field obeying Poissonian statistics, the normalized second-order correlation function vanishes ($K_{l(2)} = 0$).

For photon squeezed states whose state vector is defined as $|\alpha, z\rangle = S(z)|\alpha\rangle$, the corresponding correlation function has the form (at small values of the squeezing parameter $r_l$)
\begin{equation}
\label{QO_cor} K_{l(2)} =-\,\frac{r_l[\alpha^2_l
e^{-i\delta}+(\alpha^*_l)^2 e^{i\delta}]}
{\mid\alpha_l\mid^4-2r_l\mid\alpha_l\mid^2 [\alpha_l^2
e^{-i\delta}+(\alpha^*_l)^2 e^{i\delta}]}.
\end{equation}
In contrast to the correlation function for squeezed photon states, $K_{l(2)}$, the corresponding function for the squeezed gluon states, $K^{b}_{l(2)}$, includes, as follows from (\ref{corforgl}), the $M_2(\theta_1,\theta_2)$ which appears because the Hamiltonian (\ref{v}) of the gluon self-interaction involves a nonlinear combination of the creation and annihilation operators of gluons with different colours and vector components.

The angular dependence of the correlation function for squeezed gluon states (with colour $b=1$) that are formed at the non-perturbative stage after a lapse of $t=0.001$ is investigated graphically at the following parameter values: $\theta_{2} = 0$; $g^2 = 4\pi$ because $\alpha_s = \displaystyle\frac{g^2}{4\pi}\sim 1$; $q_{o}^{2} = 1\;\mathrm{GeV}^{2}$ that corresponds to the gluon virtuality at the
beginning of the non-perturbative stage; $k_{o} = \displaystyle\frac{\sqrt{s}}{2\langle n_{\mathrm{gluon}}\rangle}$ corresponds to a gluon energy in the case of 2-jet events; $\sqrt{s} = 91\;\mathrm{GeV}$ and $\langle n_{\mathrm{gluon}}\rangle = |\,\alpha|^{\,2} + 7|\,\beta|^{\,2}$. 

\begin{figure}[h]
     \leavevmode
\centering
\includegraphics[width=2.5in, height=1.7in, angle=0]{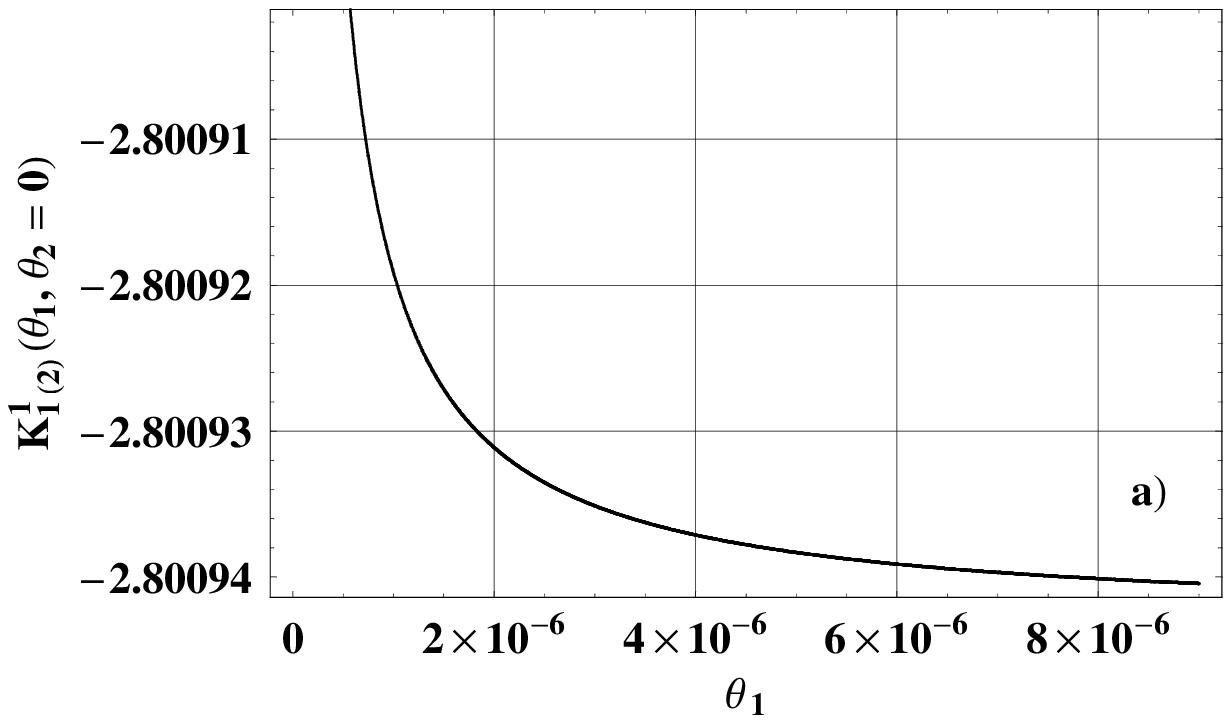}
\includegraphics[width=2.5in, height=1.7in, angle=0]{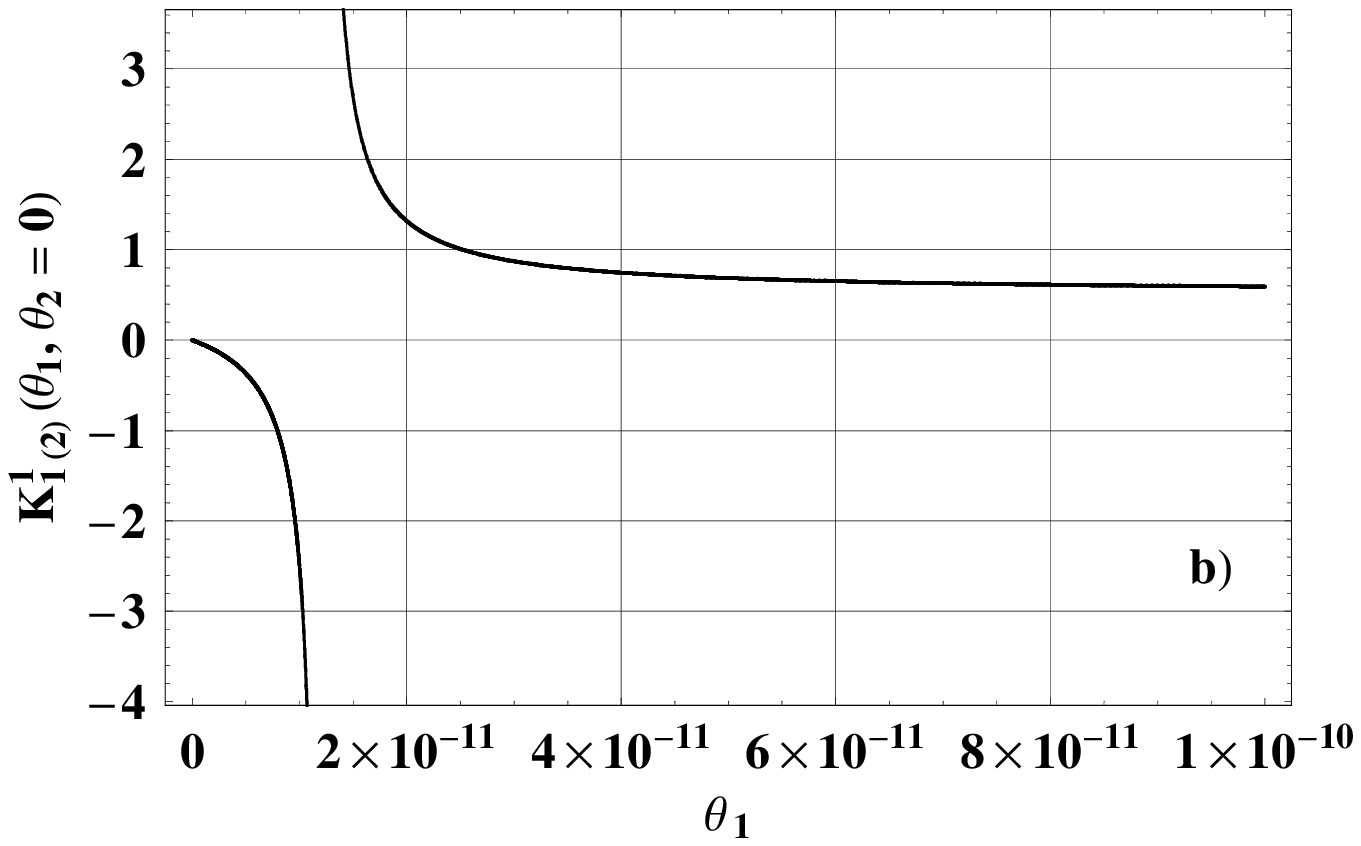}
\caption{{\small The angular dependence of the cophased ($\delta = 0$) squeezed gluon correlation function at: a)~$|\alpha|^{2}~=~1, |\beta|^2~=~1$, b) ~$|\alpha|^{2}~=~3~,~|\beta|^2~=~1$.}}
\label{firstfig}
\end{figure}

If the amplitude $|\alpha|$ of the gluon field being considered is equal to the amplitudes $|\beta|$ of the cophased gluon fields having different colours and vector components, then the values of the correlation function lie in the negative region (Fig.\ref{firstfig}a), and there occurs the antibunching of gluons with the corresponding sub-Poissonian distribution. In this case, the correlation function tends to a constant ($K^{1}_{1(2)}(\theta_1,\theta_2 = 0) = - 2.80094$) as the angle $\theta_1$ increases.  The behaviour of the angular correlations of the cophased squeezed gluon states $(\delta = 0)$ is similar to correlations of the analogous photon states at small values of the squeezing parameter \cite{QO_Japan}.
If amplitude of selected gluon field with the colour ($b = 1$)
$\alpha$ begins to dominate in relation to the amplitudes of other colour fields ($b \neq 1$), that is, $\alpha > \beta$, then correlation function involves a singularity
(Fig.\ref{firstfig}b)\, at $\theta_{1}\approx 1.208725\times 10^{-11}$ ($|\alpha|^2 = 3$, $|\beta|^2 = 1$).

For antiphased squeezed states of the gluons ($\delta = \pi$) correlation function lies in positive region and there occurs gluon bunching with the corresponding super-Poissonian distribution. In this case correlation function grows fast at small angles $\theta_1$ and tends to a constant irrespective of the values of the
amplitudes $\alpha$ and $\beta$.

\begin{figure}[h]
     \leavevmode
\centering
\includegraphics[width=3.5in, height=2in, angle=0]{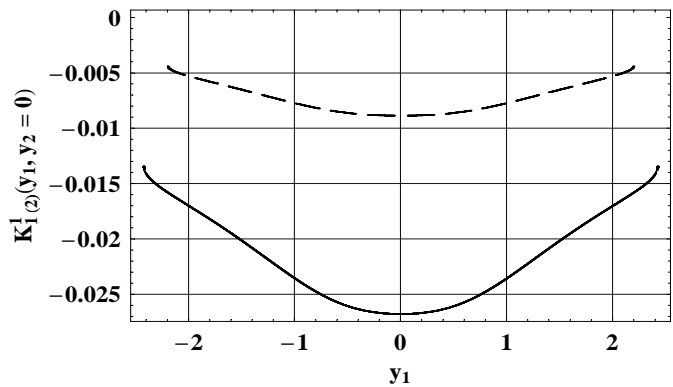}
\caption{{\small The rapidity dependence of the cophased ($\delta = 0$) gluon squeezed correlation function at $y_2 = 0$: $|\alpha|^{2}~=~1,
|\beta|^2=1 \--$ solid line, $|\alpha|^{2}=3, |\beta|^2=1
\--$ dotted line.}}
\label{second}
\end{figure}

By using the transformation
\begin{equation}
\sin\theta= \sqrt{1-\displaystyle\frac{\tanh^2 \!y}{u_1}},
\end{equation}
we can rewrite the correlation function for squeezed gluon states in terms of rapidities 
\begin{equation}
\begin{array}{l}
\label{gl_cor} K^{b}_{l(2)}(\theta_1,\theta_2) =-\,M_1 (y_1,
y_2)/\{\mid\alpha^b_l\mid^4-2\mid\alpha^b_l\mid^2M_1 (y_1, y_2)+
M_2 (y_1, y_2)\},
\end{array}
\end{equation}
Rapidity correlations of the cophased gluon squeezed states (Fig.\ref{second}) fall within the region of negative values and have a minimum in the center ($K^{b}_{l(2)}(y_1=y_2=0) = -0.0267894$ at $|\alpha|^2 = 1$). For $|\alpha| > |\beta|$ the correlation function has a less pronounced minimum in the center\\ $K^{b}_{l(2)}(y_1=y_2=0) = -0.00887147$ at $|\alpha|^2 = 3$. 

Rapidity correlations of the antiphased gluon squeezed states fall within the region of positive values and have a maximum in the center ($K^{b}_{l(2)}(y_1=y_2=0) = 0.0241966$ at $|\alpha|^2 = 1,$\\  $K^{b}_{l(2)}(y_1=y_2=0) = 0.0080179$ at $|\alpha|^2 = 3$).

\begin{figure}[h]
     \leavevmode
\centering
\includegraphics[width=2.5in, height=1.5in, angle=0]{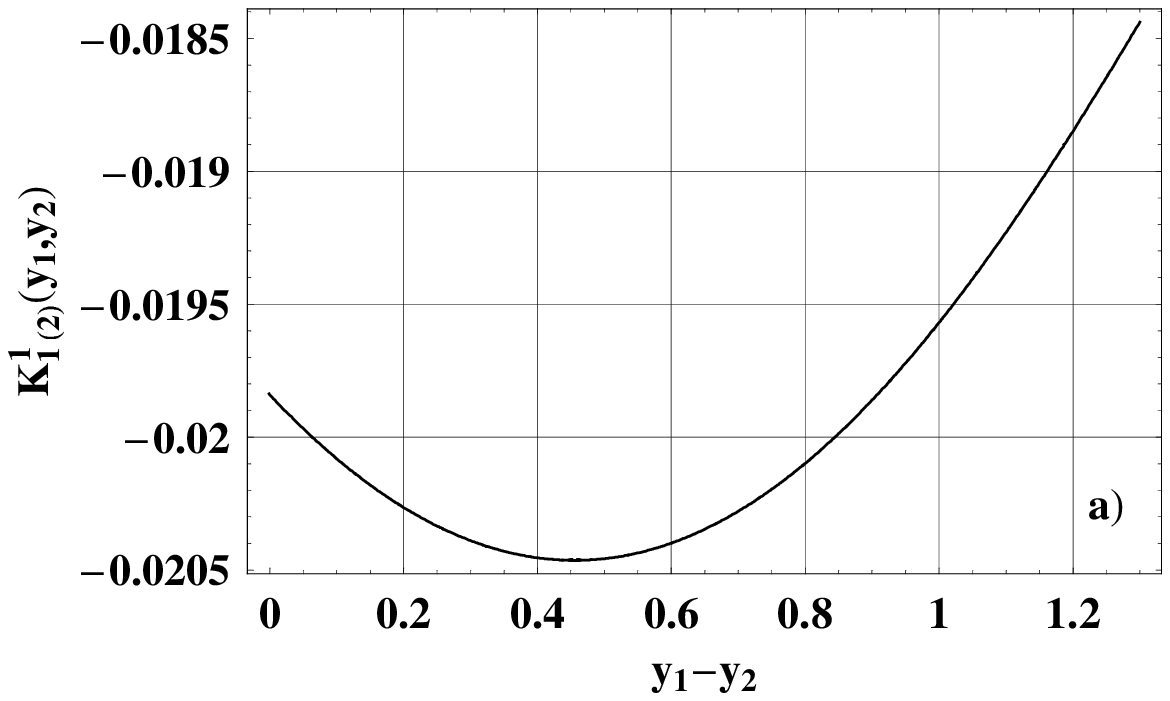}
\includegraphics[width=2.5in, height=1.5in, angle=0]{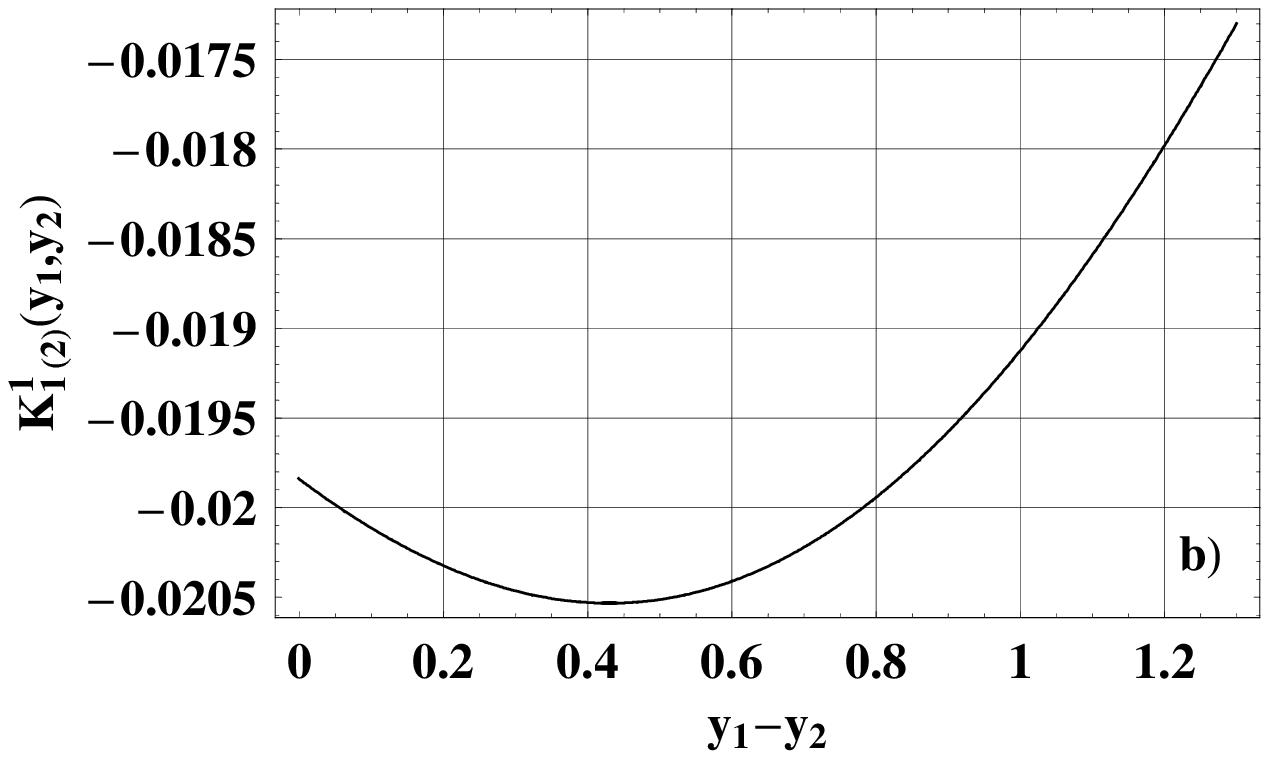}
\caption{{\small The rapidity dependence of the cophased gluon squeezed
correlation function at $\sqrt{s} = 35$ Gev $|\alpha|^{2}~=~1, |\beta|^2=1$: a) $-1\leq y_1 + y_2\leq 0, $ b) $0\leq y_1 + y_2\leq 1 $.}}
\label{fifth}
\end{figure}

It should be noted that behaviour of the cophased gluon squeezed
correlation function $K^{b}_{l(2)}(y_1,y_2)$ at $\sqrt{s} = 35$ GeV (Fig.\ref{fifth}) is similar to hadron correlations with distinctive minimum at $y_1 - y_2 = 0.45$ \cite{Kittel}.

Thus the behavior of rapidity correlations for the cophased gluon squeezed states under investigation suggests that, at the non-perturbative stage of evolution of a QCD jet, there exists the effect of the gluon antibunching with the corresponding sub-Poissonian statistics.

\section{Chaos in jets}
There are a lot of evidences that phenomenon of a chaos may exist in high-energy physics (fractality for the factorial moments \cite{Kittel}), in nuclear
physics (energy spacing distributions \cite{Zasl}), in quantum mechanics (chaos
assisted quantum tunnelling \cite{Lin}). 

There is also direct relation between chaos and squeezing in some quantum-mechanical systems \cite{Aleks1}-\cite{Aleks3}.  Let us verify this statement by the example of the next Hamiltonians:
\begin{equation}\label{H1}
  1)\qquad H_1 = \frac{p_1^2}{2} + \frac{p_2^2}{2} + \frac{g^2}{2}
q_1^2 q_2^2,
\end{equation}
\begin{equation}\label{H2}
  2)\qquad H_2 = \frac{p_1^2}{2} + \frac{p_2^2}{2} + \frac{q_1^2}{2}+ \frac{q_2^2}{2}
  + \frac{g}{2} (p_1 q_1 + q_1 p_1 + p_2 q_2 + q_2 p_2),
\end{equation}
\begin{equation}\label{H3}
  3)\qquad H_3 = g (p_1 q_2 + q_1 p_2).
\end{equation}

Hamiltonian $H_1$ is a mechanical analogue of Hamiltonian of the
Yang-Mills field for SU(2) gauge in the case of spatially
homogenous fields $(\partial_i A_j = 0)$ and $A_1^2=A_2^1=0, A_i^3 = 0$
\cite{Sav}. Hamiltonians $H_2$ and $H_3$ correspond to
Hamiltonians described degenerate and non-degenerate parametric
amplifiers in the interaction picture \cite{QO_Walls}.

The local stability of the dynamical system is determined by the
eigenvalues of the stability matrix G which has the next form
\cite{Toda}
\begin{equation}
\label{G}
 G= \left [ \begin {array}{cc}
           -\frac{\partial ^2 H}{\partial p_i \partial q_j} &
           -\frac{\partial ^2 H}{\partial q_i \partial q_j}\\ \\
            \frac{\partial ^2 H}{\partial p_i \partial p_j} &
            \frac{\partial ^2 H}{\partial q_i \partial p_j}
           \end {array} \right ]
\end{equation}

If at least one of the eigenvalues of the matrix G is real then
the separation of the trajectories grows exponentially and the
motion is unstable. Imaginary eigenvalues correspond to stable
motion. It was showed for $H_1$ \cite{Salas} that the stability
matrix G has real eigenvalues. Therefore the motion of such
systems is unstable.

The eigenvalues of G for systems described by $H_2$ and $H_3$ are
$\lambda_{1,2} = \pm \sqrt{g^2 - 1}, \lambda = g$ correspondingly,
where we take into account that the interaction constant $g$ is
real and positive. Obviously, the local nonstability for $H_2$
exists if $g^2 > 1$ and in the case $H_3$ --- for an arbitrary $g$. Thus the
motion of considered physical systems can be chaotic.

Investigating the variances $({\scriptstyle\triangle} p_i)^2$ and
$({\scriptstyle\triangle} q_i)^2$ we can show that in the case of
the Hamiltonian $H_1$ SS are not appear. At the same time from QO
it is well known that SS appear for systems described by $H_2$
and $H_3$. It should be noted, if for two-mode degenerate
parametric amplifier squeezing exists for single mode, so in the
case of non-degenerate parametric amplifier we have two-mode
squeezed states and the squeezing does not exist in the single
mode \cite{QO_Walls}.

Hamiltonians $H_2$ and $H_3$, for which the squeezing exists, have
terms proportional to the product of the canonical momentum and
coordinate $(pq)$. Indeed, let consider the next expression of the
variance $({\scriptstyle\triangle} p_i)^2$
\begin{equation}
({\scriptstyle\triangle} p_i)^2 = <\!f|p_i^2|f\!> -
<\!f|p_i|f\!>^2,
\end{equation}
where the evaluate vector $|f\!>$ is a solution of the Schr$\mathrm{\ddot
o}$dinger equation for small time
\begin{equation}
|f\!> \simeq |i\!> - \:i t H \,|i\!>
\end{equation}
Thus we have
\begin{equation}
\label{p} ({\scriptstyle\triangle} p_i)^2 = <\!i|p_i^2|i\!> -
<\!i|p_i|i\!>^2 - \:i t (<\!i|[p_i^2,H]|i\!> - \,2
<\!i|p_i|i\!><\!i|[p_i,H]|i\!>).
\end{equation}
We choose coherent state as initial state vector $|i\!> =
|\,\alpha\!>$ because it is most similar to the classical case,
any vector may be decompose on these basis vectors and these
states are widely used in QO. Then as followed from (\ref{p}) term
which is proportional to time t is a measure of departure from
initial value of the variance
\begin{equation}\label{dispers}
- \:i t (<\!\alpha|[p_i^2,H]|\alpha\!> - 2
<\!\alpha|p_i|\alpha\!><\!\alpha|[p_i,H]|\alpha\!>)= - t
<\!\alpha|[q_i,[p_i,H]]|\alpha\!>.
\end{equation}
As considered Hamiltonians are to polynomial like of $p$ and $q$
we may write (\ref{dispers}) in the next general form
\begin{equation}\label{dispers1}
- t <\!\alpha|[q_i,[p_i,H]]|\alpha\!> = - t
<\!\alpha|\frac{\partial ^2 H}{\partial p_i \partial
q_i}|\alpha\!>.
\end{equation}
Squeezed states exist when
$({\scriptstyle\triangle} p_i)^2 \neq <\!\alpha|p_i^2|\alpha\!> -
<\!\alpha|p_i|\alpha\!>^2$ that is as followed from (\ref{p}),(\ref{dispers1})
\begin{equation}\label{S_c}
<\!\alpha|\frac{\partial ^2 H}{\partial p_i \partial
q_i}|\alpha\!>\neq 0.
\end{equation}
Analogously it can be shown that the multimode SS exist when
\begin{equation}\label{S_c1}
<\!\alpha|\frac{\partial ^2 H}{\partial p_i \partial
q_j}|\alpha\!> + <\!\alpha|\frac{\partial ^2 H}{\partial q_i \partial
p_j}|\alpha\!>\neq 0,\qquad (i\neq j).
\end{equation}
This means that if components $\frac{\partial ^2 H}{\partial p_i
\partial q_i}$ or sum of the components $\frac{\partial ^2 H}{\partial p_i \partial q_j}$ and $\frac{\partial ^2 H}{\partial q_i \partial p_j}$ from stability matrix G (\ref{G}) are not equal
zero then for corresponding quantum system the squeezing exists.

One of criteria of the appearance chaos is a fast exponential
decreasing of the Green function \cite{Kuzmin} which is analogy of the
 correlator of the field operators. Therefore in order to find conditions of appearance the non-stable motion in
physical systems ($H_2$ and $H_3$) for which the squeezing exists
we investigate the next temporal correlators of the field operators
\begin{equation}\label{R}
  R_{ij} = <\!a_i^+(t)a_j(0)\!>.
\end{equation}
Here averaging is made over initial coherent states
$|\,\alpha\!>$, $a_i^+(t)$ is a solution of the Heisenberg
equations in the interaction picture.

For two-mode degenerate parametric amplifier we have
\cite{QO_Walls}
\begin{equation}
\left.
\begin{array}{c}
\displaystyle a_1^+(t) = a_1^+ \cosh (gt) + a_1 \sinh (gt), \\ \\
\displaystyle a_2^+(t) = a_2^+ \cosh (gt) + a_2 \sinh (gt).
\end{array}
\right\}
\end{equation}
Substituting these expressions in (\ref{R}) we obtain the explicit
forms of the correlators
\begin{equation}\label{R1}
R_{ij}(t)=\frac{\alpha_j}{2}(\alpha_i^* + \alpha_i)e^{gt} +
\frac{\alpha_j}{2}(\alpha_i^* - \alpha_i)e^{-gt}, \qquad i,j=1,2.
\end{equation}
 As follows from (\ref{R1}) the correlators $R_{ij}(t)$ decrease exponentially at
equal phases $\varphi_1~\!\!=~\!\!\varphi_2~\!\!=~\!\!\displaystyle\frac{\pi}{2},\!
\displaystyle\frac{3\pi}{2}\;$ 
$(\alpha_i = |\alpha_i|\,e^{i\varphi_i}).$   In this case we have
\begin{equation}\label{R11}
R_{ij}(t)=|\alpha_i| |\alpha_j| e^{-gt}, \qquad i,j=1,2.
\end{equation}
As squeezing does not depend on $\alpha_1$ and $\alpha_2$
\cite{QO_Walls} then in this case the effects of squeezing and
chaos coexist.

Using the solutions of the evolutions equations for $a_1^+(t)$ and
$a_2^+(t)$ in the case non-degenerate parametric amplifier
\cite{QO_Walls} we can obtain explicit expression for
corresponding correlators
\begin{equation}
\left.
\begin{array}{c}
\displaystyle R_{ii}(t)=\frac{1}{2}(|\alpha_i|^2 +
\alpha_1\alpha_2)e^{gt} + \frac{1}{2}(|\alpha_i|^2 -
\alpha_1\alpha_2)e^{-gt}, \qquad
i=1,2, \\ \\
\displaystyle R_{ij}(t)=\frac{\alpha_j}{2}(\alpha_i^* +
\alpha_j)e^{gt} + \frac{\alpha_j}{2}(\alpha_i^* - \alpha_j)e^{-gt},
\qquad (i\neq j),\quad i,j=1,2.
\end{array}
\right\}
\end{equation}
Obviously, these correlators are exponentially reduced at
$|\alpha_1|= |\alpha_2|\equiv |\alpha|, \varphi_1 = \varphi_2 = \displaystyle\frac{\pi}{2}, \displaystyle\frac{3\pi}{2}$
and then $R_{ii}(t) = R_{ij}(t),$
\begin{equation}\label{R2}
R_{ij}(t)=|\alpha|^2 e^{-gt}.
\end{equation}
This clearly demonstrates the possibility of
existence of chaos when the squeezing condition is fulfilled for
any $\alpha$.

The problem of the existence of a chaos in model SU(2)-jet merits especial attention. Here we take the case of SU(2)-group only for simplification of calculations.
Respective Hamiltonian  of the interaction $V_\mathrm{int}$ is obtained from the Hamiltonian for SU(3)-jet (\ref{v}) by replacing structure constants of the SU(3)-group the corresponding SU(2)-group constant, that is $ f_{abc} \rightarrow \varepsilon_{abc}\; (a,b,c = \overline{1,3}),$ and take into account that $\varepsilon_{abc}\varepsilon_{ade} = \delta_{bd}\delta_{ce} - \delta_{be}\delta_{cd}$

\begin{equation}
\begin{array}{l}
\label{vsu(2)} V_\mathrm{int}=\displaystyle\frac{k_0^4}{4(2\pi)^3}\left
(\!1-\frac{q_0^2}{k_0^2}\!\right )^{\!3/2}\!\!g^2\, \pi\,
  \Biggl\{ \!\left (\!2-\displaystyle\frac{q_0^2}{k^2_0}-\frac{\sin^2\theta}2
  \Bigl (1-\displaystyle\frac{q_0^2}{k^2_0} \Bigr )\right)\! \Bigl[a^{bcbc}_{1212}+a^{bcbc}_{1313}-a^{bccb}_{1212}-a^{bccb}_{1313}\Bigr]\\
\qquad\qquad + \left (\!1+\sin^2\theta
  \Bigl (1-\displaystyle\frac{q_0^2}{k^2_0} \Bigr )\right) \Bigl[a^{bcbc}_{2323}-a^{bccb}_{2323}\Bigr]
\Biggr\}\sin\theta\:d\theta.
\end{array}
\end{equation}

The local instability for the Hamiltonian (\ref{vsu(2)}) has been verified by Toda criterion. Analysis was made numerically accordingly the next algorithm:
\begin{description}
  \item[1.] we come to classical Hamiltonian by keeping the order of
operators $a^{+},a$ and consider them as c-numbers ($\alpha^*, \alpha$);
  \item[2.] we have 18 variables and calculate the instability matrix  $18\times 18$
for this case;
  \item[3.] next step is calculation of its eigenvalues to find out whether
they are real or imaginary.
\end{description}
As a result the following conclusions have been obtained:\\
1) If all variable $\alpha$ and $\alpha^{*}$ are real or imaginary then the system of gluons described by the above mentioned
Hamiltonian (\ref{vsu(2)}) is strictly ordered and effect of the squeezing is absent.

\noindent
2) If at least one of $\alpha$ or $\alpha^{*}$ is imaginary and other are
real or at least one of $\alpha$ and $\alpha^{*}$ is real  and other are
imaginary --- we have the local instability, which can lead to chaotical system.

\section{Conclusion}
Investigating of the gluon fluctuations we have proved theoretically the possibility of existence of the gluon single-mode SS at non-perturbative stage of the QCD jet evolution. The emergence of such remarkable states becomes possible owing to the self-interaction of gluons with different colour indices.

As one of identification criterion of existence of gluon SS can
served correlation function. Therefore we have analyzed the behaviour of angular and rapidity correlations and have compared our results with the corresponding correlation function for photon squeezed states, which was comprehensively investigated in quantum optics. The form of the normalized correlation function $K^{b}_{l(2)}$ for cophased squeezed states specifies the gluon antibunching effect if the amplitudes of all gluon fields (with various colour and vector components) are equal to one another. Such behaviour of angular correlations is analogous to the behaviour of the corresponding correlations of the photon squeezed states at small values of the squeezing parameter. At the same time, there is distinction between them: in contrast to the normalized correlation function known in quantum optics, the correlation function of the gluon SS has a singularity if the amplitude for the fixed-colour gluon field being studied is greater than the amplitudes for gluon fields with other colour indices. The correlations of the cophased squeezed states specifies the presence of the gluon antibunching effect, whereas the gluon bunching occurs for antiphased squeezed states. Hence, non-perturbative gluon evolution makes a contribution to the parton distribution prepared by the perturbative stage of jet evolution in the form of a sub-Poissonian (cophased squeezed states) or a super-Poissonian (antiphased squeezed states) distributions.

Thus, the behaviour of the two-particle angular and rapidity correlations can serve as one of the criteria of the existence of squeezed gluon states. At the same time for a comparison of our results with experimental data, we must take into account the contribution of the perturbative stage of jet evolution and hadronization effects. This can be done by using Monte-Karlo methods and will be the subject of our further investigations.

In this paper we investigate the possibility of coexistence both
condition of squeezing and chaos for some physical systems:
mechanical model of Yang-Mills field for SU(2) gauge, two-mode
degenerate and non-degenerate parametric amplifiers, SU(2)-jet model. Using Toda
criterion we check the local instability in the corresponding 
classical systems and determine conditions of the coexistence of this
effect and squeezing at small time. It was shown that the squeezing 
exists at small time under condition that either components $\frac{\partial ^2 H}{\partial p_i
\partial q_i}$ or sum of the components $\frac{\partial ^2 H}{\partial p_i \partial q_j}$ and $\frac{\partial ^2 H}{\partial q_i \partial p_j}$ from stability matrix G are not equal zero.
 
The possibility of existence of squeezing and chaos in
corresponding quantum systems is studied by using correlators of the field
operators. The correlators of the field operators
for degenerate and non-degenerate parametric amplifiers decrease
exponentially independently from amplitude values of the initial 
coherent fields $|\alpha_1|, |\alpha_2|$ if phases $\varphi_1$ and $\varphi_2$ 
coincide and equal $\displaystyle\frac{\pi}{2}$ or $\displaystyle\frac{3\pi}{2}.$ Therefore in this case chaos may exist. Moreover for these systems squeezing does not depend on value of amplitude of initial coherent field. 

Under investigation the local instability within SU(2)-jet model it was numerically shown that this effect existences under condition if at least one of amplitude of the coherent fields $\alpha$ or $\alpha^{*}$ is imaginary and other are
real and vice versa.

Thus for  degenerate and non-degenerate parametric amplifiers and also for  SU(2)-jet model the effects of the squeezing and chaos can coexist under some conditions.

\section*{Appendix A}
\begin{equation*}
\begin{split}
\prod\limits_{c=1}^8\!\prod\limits_{l=1}^3\!\!\mid\!
\alpha^c_l(t)\rangle &\simeq (1-2it\pi u_3 \sin\theta d\theta)|\,i\,\rangle - 2it\pi u_4 \sin\theta d\theta \sum\limits_{k=1}^8 \alpha_1^k a^{k+}_1 |\,i\,\rangle - 2it\pi u_5 \sin\theta d\theta \sum\limits_{k=1}^8 \sum\limits_{m=2}^3 \alpha_m^k a^{k+}_m |\,i\,\rangle \\&  - 2it\pi u_2 \sin\theta d\theta\{(1+u_1)[B_{1122} + B_{2211} + B_{1133} + B_{3311} - B_{1212} - B_{2121} - B_{1313} - B_{3131}]\\& + [B_{2233}+B_{3322}-B_{2323}-B_{3232}] + u_1 \sin^2 \theta \, [B_{2233}+B_{3322}-B_{2323}-B_{3232}\\& - \frac12(B_{1122} + B_{2211} + B_{1133} + B_{3311} - B_{1212} - B_{2121} - B_{1313} - B_{3131})]\\& + 2(1+u_1)[C_{1122} + C_{2211} + C_{1133} + C_{3311} + D_{1212} + D_{2121} + D_{1313} + D_{3131}]\\& + 2 [C_{2233} + C_{3322} + D_{2323} + D_{3232}] + 2 u_1 \sin^2 \theta\,[C_{2233} + C_{3322} + D_{2323} + D_{3232}\\& - \frac12 (C_{1122} + C_{2211} + C_{1133} + D_{1212} + D_{2121} + D_{1313} + D_{3131})]\} |\,i\,\rangle, \qquad\qquad\qquad\quad (A.1)
\end{split}
\end{equation*}
where $
|\,i\,\rangle = \prod\limits_{c=1}^8\!\prod\limits_{l=1}^3\!\!\mid\!
\alpha^c_l(0)\rangle,
$ $u_5 = \displaystyle\frac{k_0^3}{2} \sqrt{u_1}\left\{\!1+u_1-\displaystyle\frac12\,\Bigl(1+u_1-\displaystyle\frac{q_0^4}{k_0^4}
\Bigr)\,
\sin^2\theta\!\right\} + 6 u_2\left\{\!2+u_1+\displaystyle\frac12\, u_1 \sin^2\theta \!\right\}\!,$

\begin{equation*}
\begin{split}
B_{ij\,mn}& = A_{mn}^{11}\mathop{{\sum}'}\limits_{k=2}^{7}\alpha_{ij}^{kk} + A_{mn}^{22}\mathop{{\sum}'}\limits_{\substack{k=1\\k\neq 2}}^{7}\alpha_{ij}^{kk} + A_{mn}^{33}\mathop{{\sum}'}\limits_{\substack{k=1\\k\neq 3}}^{7}\alpha_{ij}^{kk} + \frac14 A_{mn}^{44}\biggl[\sum\limits_{\substack{k=1\\k\neq 4}}^{7}\alpha_{ij}^{kk} + 3 \sum\limits_{k=5,\,8}\alpha_{ij}^{kk} + \sqrt{3}(\alpha_{ij}^{38}+\alpha_{ij}^{83})\biggr]\\&
+ \frac14 A_{mn}^{55}\biggl[\sum\limits_{\substack{k=1\\k\neq 5}}^{7}\alpha_{ij}^{kk} + 3 \sum\limits_{k=4,\,8}\alpha_{ij}^{kk} + \sqrt{3}(\alpha_{ij}^{38}+\alpha_{ij}^{83})\!\biggr] + \frac14 A_{mn}^{66}\biggl[\sum\limits_{\substack{k=1\\k\neq 6}}^{7}\alpha_{ij}^{kk} + 3 \sum\limits_{k=7,\,8}\alpha_{ij}^{kk} - \sqrt{3}(\alpha_{ij}^{38}+\alpha_{ij}^{83})\!\biggr]\\&
+ \frac14 A_{mn}^{77}\biggl[\sum\limits_{k=1}^{6}\alpha_{ij}^{kk} + 3 \sum\limits_{k=6,\,8}\alpha_{ij}^{kk} - \sqrt{3}(\alpha_{ij}^{38}+\alpha_{ij}^{83})\!\biggr] + \frac34\, A_{mn}^{88}\sum\limits_{k=4}^{7}\alpha_{ij}^{kk}, \qquad\qquad\qquad\qquad\qquad (A.2)
\end{split}
\end{equation*}
\begin{equation*}
\begin{split}
C_{ij\,mn}&= A_{mn}^{12}\Bigl\{- \alpha_{ij}^{21} + \frac14 \bigl[\alpha_{ij}^{76} - \alpha_{ij}^{67} + \alpha_{ij}^{45} - \alpha_{ij}^{54}\bigr]\Bigr\} + A_{mn}^{13}\Bigl\{- \alpha_{ij}^{31} + \frac14 \bigl[\alpha_{ij}^{64} - \alpha_{ij}^{46} + \alpha_{ij}^{75} - \alpha_{ij}^{57}\bigr]\Bigr\}\\&
+\frac12\,A_{mn}^{14}\Bigl\{\alpha_{ij}^{25} - \alpha_{ij}^{36}  + \frac12 \bigl[\alpha_{ij}^{52} - \alpha_{ij}^{63} - \alpha_{ij}^{41} - \sqrt{3}\,\alpha_{ij}^{68}\bigr]\!\Bigr\}\\&
-\frac12\,A_{mn}^{15}\Bigl\{\alpha_{ij}^{24} + \alpha_{ij}^{37}  + \frac12 \bigl[\alpha_{ij}^{42} + \alpha_{ij}^{73} + \alpha_{ij}^{51} + \sqrt{3}\,\alpha_{ij}^{78}\bigr]\!\Bigr\}\\&
+\frac12\,A_{mn}^{16}\Bigl\{\alpha_{ij}^{34} - \alpha_{ij}^{27}  + \frac12 \bigl[\alpha_{ij}^{43} - \alpha_{ij}^{72} - \alpha_{ij}^{61} - \sqrt{3}\,\alpha_{ij}^{48}\bigr]\!\Bigr\}\\&
+\frac12\,A_{mn}^{17}\Bigl\{\alpha_{ij}^{35} + \alpha_{ij}^{26}  + \frac12 \bigl[\alpha_{ij}^{53} + \alpha_{ij}^{62} - \alpha_{ij}^{71} - \sqrt{3}\,\alpha_{ij}^{58}\bigr]\!\Bigr\} 
+ \frac{\sqrt{3}}{4}\,A_{mn}^{18}\Bigl\{\alpha_{ij}^{46} + \alpha_{ij}^{64} + \alpha_{ij}^{57} + \alpha_{ij}^{75}\!\Bigr\}\\&
+ A_{mn}^{23}\Bigl\{- \alpha_{ij}^{32} + \frac14 \bigl[\alpha_{ij}^{65} - \alpha_{ij}^{56} + \alpha_{ij}^{47} - \alpha_{ij}^{74}\bigr]\Bigr\}
+ \frac12\,A_{mn}^{24}\Bigl\{\alpha_{ij}^{37} - \alpha_{ij}^{15}  + \frac12 \bigl[\alpha_{ij}^{73} - \alpha_{ij}^{42} - \alpha_{ij}^{51} + \sqrt{3}\,\alpha_{ij}^{78}\bigr]\!\Bigr\}\\&
+ \frac12\,A_{mn}^{25}\Bigl\{\alpha_{ij}^{14} - \alpha_{ij}^{36}  + \frac12 \bigl[\alpha_{ij}^{41} - \alpha_{ij}^{63} - \alpha_{ij}^{52} - \sqrt{3}\,\alpha_{ij}^{68}\bigr]\!\Bigr\}\\&
+ \frac12\,A_{mn}^{26}\Bigl\{\alpha_{ij}^{35} + \alpha_{ij}^{17}  + \frac12 \bigl[\alpha_{ij}^{53} + \alpha_{ij}^{71} - \alpha_{ij}^{62} - \sqrt{3}\,\alpha_{ij}^{58}\bigr]\!\Bigr\}\\&
- \frac12\,A_{mn}^{27}\Bigl\{\alpha_{ij}^{16} + \alpha_{ij}^{34}  + \frac12 \bigl[\alpha_{ij}^{72} + \alpha_{ij}^{43} + \alpha_{ij}^{61} - \sqrt{3}\,\alpha_{ij}^{48}\bigr]\!\Bigr\}
+ \frac{\sqrt{3}}{4}\,A_{mn}^{28}\Bigl\{\alpha_{ij}^{56} + \alpha_{ij}^{65} - \alpha_{ij}^{47} - \alpha_{ij}^{74}\!\Bigr\}\\&
+ \frac12\,A_{mn}^{34}\Bigl\{\alpha_{ij}^{16} - \alpha_{ij}^{27}  + \frac12 \bigl[\alpha_{ij}^{61} - \alpha_{ij}^{43} - \alpha_{ij}^{72} - \sqrt{3}\,\alpha_{ij}^{48}\bigr]\!\Bigr\}\\&
+ \frac12\,A_{mn}^{35}\Bigl\{\alpha_{ij}^{26} + \alpha_{ij}^{17}  + \frac12 \bigl[\alpha_{ij}^{71} + \alpha_{ij}^{62} - \alpha_{ij}^{53} - \sqrt{3}\,\alpha_{ij}^{58}\bigr]\!\Bigr\}\\&
- \frac12\,A_{mn}^{36}\Bigl\{\alpha_{ij}^{25} + \alpha_{ij}^{14}  + \frac12 \bigl[\alpha_{ij}^{52} + \alpha_{ij}^{41} + \alpha_{ij}^{63} - \sqrt{3}\,\alpha_{ij}^{68}\bigr]\!\Bigr\}\\&
+ \frac12\,A_{mn}^{37}\Bigl\{\alpha_{ij}^{24} - \alpha_{ij}^{15}  + \frac12 \bigl[\alpha_{ij}^{42} - \alpha_{ij}^{51} - \alpha_{ij}^{73} + \sqrt{3}\,\alpha_{ij}^{78}\bigr]\!\Bigr\}+ \frac{\sqrt{3}}{4}\,A_{mn}^{38}\Bigl\{\alpha_{ij}^{44} + \alpha_{ij}^{55} - \alpha_{ij}^{66} - \alpha_{ij}^{77}\!\Bigr\}\\&
+ \frac{1}{4}\,A_{mn}^{45}\Bigl\{\alpha_{ij}^{12} - \alpha_{ij}^{21} + \alpha_{ij}^{67} - \alpha_{ij}^{76} - 4\, \alpha_{ij}^{54}\!\Bigr\}\\&
+ \frac{1}{4}\,A_{mn}^{46}\Bigl\{\alpha_{ij}^{31} - \alpha_{ij}^{13} + \alpha_{ij}^{75} - \alpha_{ij}^{64} + 2 \alpha_{ij}^{57} + \sqrt{3}\, (\alpha_{ij}^{18} + \alpha_{ij}^{81})\!\Bigr\}\\&
+ \frac{1}{4}\,A_{mn}^{47}\Bigl\{\alpha_{ij}^{23} - \alpha_{ij}^{32} - \alpha_{ij}^{65} - \alpha_{ij}^{74} - 2 \alpha_{ij}^{56} - \sqrt{3}\, (\alpha_{ij}^{28} + \alpha_{ij}^{82})\!\Bigr\}- \frac{\sqrt{3}}{4}\,A_{mn}^{48}\Bigl\{\alpha_{ij}^{34} + \alpha_{ij}^{16} - \alpha_{ij}^{27} + \sqrt{3}\,\alpha_{ij}^{84}\!\Bigr\}\\&
+ \frac{1}{4}\,A_{mn}^{56}\Bigl\{\alpha_{ij}^{32} - \alpha_{ij}^{74} - \alpha_{ij}^{23} - \alpha_{ij}^{65} - 2 \alpha_{ij}^{47} + \sqrt{3}\, (\alpha_{ij}^{28} + \alpha_{ij}^{82})\!\Bigr\}\\&
+ \frac{1}{4}\,A_{mn}^{57}\Bigl\{\alpha_{ij}^{64} + \alpha_{ij}^{31} - \alpha_{ij}^{13} - \alpha_{ij}^{75} + 2 \alpha_{ij}^{46} + \sqrt{3}\, (\alpha_{ij}^{18} + \alpha_{ij}^{81})\!\Bigr\}- \frac{\sqrt{3}}{4}\,A_{mn}^{58}\Bigl\{\alpha_{ij}^{17} + \alpha_{ij}^{26} + \alpha_{ij}^{35} + \sqrt{3}\,\alpha_{ij}^{85}\!\Bigr\}\\&
+ \frac{1}{4}\,A_{mn}^{67}\Bigl\{\alpha_{ij}^{45} - \alpha_{ij}^{54} + \alpha_{ij}^{21} - \alpha_{ij}^{12} - 4\, \alpha_{ij}^{76}\!\Bigr\}- \frac{\sqrt{3}}{4}\,A_{mn}^{68}\Bigl\{\alpha_{ij}^{14} + \alpha_{ij}^{25} - \alpha_{ij}^{36} + \sqrt{3}\,\alpha_{ij}^{86}\!\Bigr\}\\&
- \frac{\sqrt{3}}{4}\,A_{mn}^{78}\Bigl\{\alpha_{ij}^{15} - \alpha_{ij}^{24} - \alpha_{ij}^{37} + \sqrt{3}\,\alpha_{ij}^{87}\!\Bigr\}, \qquad\qquad\qquad\qquad\qquad\qquad\qquad\qquad\qquad\qquad\qquad (A.3)
\end{split}
\end{equation*}
\begin{equation*}
\begin{split}
D_{ij\,mn}&= A_{mn}^{12}\Bigl\{2 \alpha_{ij}^{12} - \alpha_{ij}^{21} + \frac34 \bigl[\alpha_{ij}^{45} - \alpha_{ij}^{54} + \alpha_{ij}^{76} - \alpha_{ij}^{67}\bigr] \!\Bigr\} + A_{mn}^{13}\Bigl\{2 \alpha_{ij}^{13} - \alpha_{ij}^{31} + \frac34 \bigl[\alpha_{ij}^{64} - \alpha_{ij}^{46} + \alpha_{ij}^{75} - \alpha_{ij}^{57}\bigr] \!\Bigr\}\\&
+ \frac14 A_{mn}^{14}\Bigl\{3\,(\alpha_{ij}^{63} - \alpha_{ij}^{52}) + \sqrt{3}\,(2 \alpha_{ij}^{86} - \alpha_{ij}^{68}) - \alpha_{ij}^{41} + 2 \alpha_{ij}^{14} \!\Bigr\}\\&
+ \frac14 A_{mn}^{15}\Bigl\{3\,(\alpha_{ij}^{42} + \alpha_{ij}^{73}) + \sqrt{3}\,(2 \alpha_{ij}^{87} - \alpha_{ij}^{78}) - \alpha_{ij}^{51} + 2 \alpha_{ij}^{15} \!\Bigr\}\\&
+ \frac14 A_{mn}^{16}\Bigl\{3\,(\alpha_{ij}^{72} - \alpha_{ij}^{43}) + \sqrt{3}\,(2 \alpha_{ij}^{84} - \alpha_{ij}^{48}) - \alpha_{ij}^{61} + 2 \alpha_{ij}^{16}\!\Bigr\}\\&
- \frac14 A_{mn}^{17}\Bigl\{3\,(\alpha_{ij}^{53} + \alpha_{ij}^{62}) - \sqrt{3}\,(2 \alpha_{ij}^{85} - \alpha_{ij}^{58}) + \alpha_{ij}^{71} - 2 \alpha_{ij}^{17}\!\Bigr\} -  \frac{\sqrt{3}}{4}\,A_{mn}^{18}\Bigl\{\alpha_{ij}^{46} + \alpha_{ij}^{64} + \alpha_{ij}^{57} + \alpha_{ij}^{75}\!\Bigr\}\\&
+ A_{mn}^{23}\Bigl\{2 \alpha_{ij}^{23} - \alpha_{ij}^{32} + \frac34 \bigl[\alpha_{ij}^{56} - \alpha_{ij}^{65} + \alpha_{ij}^{74} - \alpha_{ij}^{47}\bigr] \!\Bigr\}\\&
+ \frac14 A_{mn}^{24}\Bigl\{3\,(\alpha_{ij}^{51} - \alpha_{ij}^{73}) - \sqrt{3}\,(2 \alpha_{ij}^{87} - \alpha_{ij}^{78}) - \alpha_{ij}^{42} + 2 \alpha_{ij}^{24}\!\Bigr\}\\&
+ \frac14 A_{mn}^{25}\Bigl\{3\,(\alpha_{ij}^{63} - \alpha_{ij}^{41}) + \sqrt{3}\,(2 \alpha_{ij}^{86} - \alpha_{ij}^{68}) - \alpha_{ij}^{52} + 2 \alpha_{ij}^{25}\!\Bigr\}\\&
- \frac14 A_{mn}^{26}\Bigl\{3\,(\alpha_{ij}^{53} + \alpha_{ij}^{71}) - \sqrt{3}\,(2 \alpha_{ij}^{85} - \alpha_{ij}^{58}) + \alpha_{ij}^{62} - 2 \alpha_{ij}^{26}\!\Bigr\}\\&
+ \frac14 A_{mn}^{27}\Bigl\{3\,(\alpha_{ij}^{43} + \alpha_{ij}^{61}) - \sqrt{3}\,(2 \alpha_{ij}^{84} - \alpha_{ij}^{48}) - \alpha_{ij}^{72} + 2 \alpha_{ij}^{27}\!\Bigr\} - \frac{\sqrt{3}}{4}\,A_{mn}^{28}\Bigl\{\alpha_{ij}^{56} + \alpha_{ij}^{65} - \alpha_{ij}^{47} - \alpha_{ij}^{74}\!\Bigr\}\\&
+ \frac14 A_{mn}^{34}\Bigl\{3\,(\alpha_{ij}^{72} - \alpha_{ij}^{61}) + \sqrt{3}\,(2 \alpha_{ij}^{84} - \alpha_{ij}^{48}) - \alpha_{ij}^{43} + 2 \alpha_{ij}^{34}\!\Bigr\}\\&
- \frac14 A_{mn}^{35}\Bigl\{3\,(\alpha_{ij}^{62} + \alpha_{ij}^{71}) - \sqrt{3}\,(2 \alpha_{ij}^{85} - \alpha_{ij}^{58}) + \alpha_{ij}^{53} - 2 \alpha_{ij}^{35}\!\Bigr\}\\&
+ \frac14 A_{mn}^{36}\Bigl\{3\,(\alpha_{ij}^{52} + \alpha_{ij}^{41}) - \sqrt{3}\,(2 \alpha_{ij}^{86} - \alpha_{ij}^{68}) - \alpha_{ij}^{63} + 2 \alpha_{ij}^{36}\!\Bigr\}\\&
+ \frac14 A_{mn}^{37}\Bigl\{3\,(\alpha_{ij}^{51} - \alpha_{ij}^{42}) - \sqrt{3}\,(2 \alpha_{ij}^{87} - \alpha_{ij}^{78}) - \alpha_{ij}^{73} + 2 \alpha_{ij}^{37}\!\Bigr\}-
\frac{\sqrt{3}}{4}\,A_{mn}^{38}\Bigl\{\alpha_{ij}^{44} + \alpha_{ij}^{55} - \alpha_{ij}^{66} - \alpha_{ij}^{77}\!\Bigr\}\\&
+ A_{mn}^{45}\Bigl\{2 \alpha_{ij}^{45} - \alpha_{ij}^{54} + \frac34 \bigl[\alpha_{ij}^{12} - \alpha_{ij}^{21} + \alpha_{ij}^{76} - \alpha_{ij}^{67}\bigr] \!\Bigr\}\\&
+ \frac14 A_{mn}^{46}\Bigl\{3\,(\alpha_{ij}^{31} - \alpha_{ij}^{13} - \alpha_{ij}^{75}) - \sqrt{3}\,(\alpha_{ij}^{18} + \alpha_{ij}^{81}) - \alpha_{ij}^{64} + 2 \alpha_{ij}^{46}\!\Bigr\}\\&
+ \frac14 A_{mn}^{47}\Bigl\{3\,(\alpha_{ij}^{23} - \alpha_{ij}^{32} + \alpha_{ij}^{65}) + \sqrt{3}\,(\alpha_{ij}^{28} + \alpha_{ij}^{82}) - \alpha_{ij}^{74} + 2 \alpha_{ij}^{47}\!\Bigr\}\\&
+\frac{\sqrt{3}}4 A_{mn}^{48}\Bigl\{2\,(\alpha_{ij}^{61} - \alpha_{ij}^{72} + \alpha_{ij}^{43}) - \alpha_{ij}^{16} + \alpha_{ij}^{27} - \alpha_{ij}^{34} + \sqrt{3}\,(2 \alpha_{ij}^{48} - \alpha_{ij}^{84})\!\Bigr\}\\&
+ \frac14 A_{mn}^{56}\Bigl\{3\,(\alpha_{ij}^{32} - \alpha_{ij}^{23} + \alpha_{ij}^{74}) - \sqrt{3}\,(\alpha_{ij}^{28} + \alpha_{ij}^{82}) - \alpha_{ij}^{65} + 2 \alpha_{ij}^{56}\!\Bigr\}\\&
+ \frac14 A_{mn}^{57}\Bigl\{3\,(\alpha_{ij}^{31} - \alpha_{ij}^{13} - \alpha_{ij}^{64}) - \sqrt{3}\,(\alpha_{ij}^{18} + \alpha_{ij}^{81}) - \alpha_{ij}^{75} + 2 \alpha_{ij}^{57}\!\Bigr\}\\&
+\frac{\sqrt{3}}4 A_{mn}^{58 }\Bigl\{2\,(\alpha_{ij}^{71} + \alpha_{ij}^{62} + \alpha_{ij}^{53}) - \alpha_{ij}^{17} - \alpha_{ij}^{26} - \alpha_{ij}^{35} + \sqrt{3}\,(2 \alpha_{ij}^{58} - \alpha_{ij}^{85})\!\Bigr\}\\&
+ A_{mn}^{67}\Bigl\{2 \alpha_{ij}^{67} - \alpha_{ij}^{76} + \frac34 \bigl[\alpha_{ij}^{21} - \alpha_{ij}^{12} - \alpha_{ij}^{45} + \alpha_{ij}^{54}\bigr] \!\Bigr\}\\&
+\frac{\sqrt{3}}4 A_{mn}^{68}\Bigl\{2\,(\alpha_{ij}^{41} + \alpha_{ij}^{52} - \alpha_{ij}^{63}) - \alpha_{ij}^{14} - \alpha_{ij}^{25} + \alpha_{ij}^{36} + \sqrt{3}\,(2 \alpha_{ij}^{68} - \alpha_{ij}^{86})\!\Bigr\}\\&
+\frac{\sqrt{3}}4 A_{mn}^{78}\Bigl\{2\,(\alpha_{ij}^{51} - \alpha_{ij}^{42} - \alpha_{ij}^{73}) - \alpha_{ij}^{15} + \alpha_{ij}^{24} + \alpha_{ij}^{37} + \sqrt{3}\,(2 \alpha_{ij}^{78} - \alpha_{ij}^{87})\!\Bigr\}, \qquad\qquad\qquad\qquad (A.4)
\end{split}
\end{equation*}
$A_{mn}^{bc} = a_m^{b+} a_n^{c+}, \qquad \alpha_{ij}^{bc} = \alpha_i^b \alpha_j^c\,.$


\begin{thebibliography}{99}
\bibitem{PQCD-Coher}
Yu. L. Dokshitzer, \textit{et al.,} Rev. Mod. Phys. \textbf{60},
373 (1988).

\bibitem{PQCD}
Yu. L. Dokshitzer, V. A. Khoze, A. H. Mueller, S. I. Troyan,
 \textit{Basics of Perturbative QCD}
(Fronti$\mathrm{\grave{e}}$res,France, 1991).

\bibitem{Kuvsh}
V. I. Kuvshinov, Acta Phys. Pol. B \textbf{10}, 19 (1979).

\bibitem{Kokoulina}
E. S. Kokoulina, V. I. Kuvshinov, Acta Phys. Pol. B \textbf{13},
553 (1982).

\bibitem{Malasa}
E. Malasa, B. Webber, Nucl. Phys. B \textbf{267}, 702 (1986).

\bibitem{Dremin}
I. M. Dremin, R. C. Hwa, Phys. Rev. D \textbf{49}, 5805 (1994).

\bibitem{Soft}
S. Lupia, W. Ochs, J. Wosiek, Nucl. Phys. B \textbf{540}, 405
(1999).

\bibitem{Nasa}
S. Ya. Kilin, V. I. Kuvshinov, S. A. Firago, Proceed. of a Workshop on Squeezed states and Uncertainty relations, (Moscow, Russia, 1992), NASA, 301 (1993).

\bibitem{acta}
V. I. Kuvshinov, V. A. Shaporov, Acta Phys. Pol. B \textbf{30}, 59
(1999).

\bibitem{NPCS}
V. I. Kuvshinov, V. A. Shaparau, Nonlinear Phenomena in Complex
Systems \textbf{3}, 28 (2000).

\bibitem{QO_Japan}
O. Hirota, \textit{Squeezed light} (Japan, Tokyo, 1992).

\bibitem{QO_Walls}
D. F. Walls, G. J. Milburn, \textit{Quantum Optics}
(Springer-Verlag, NY., USA, 1995).

\bibitem{QO_Cambr}
M. O. Scully, M. S. Zubairy M.S, \textit{Quantum Optics}
(Cambridge University Press, 1997).

\bibitem{Kilin}
S.Ya. Kilin, \textit{Quantum Optics}  (Minsk, 1990). (in Russian)

\bibitem{PSS}
I. M. Dremin, Phys. Lett. A \textbf{193}, 209 (1994).

\bibitem{UA5}
G.J. Alner et al., (UA5 coll.) Phys. Rep. {\bf 154}, 247 (1987)

\bibitem{DELPHI}
P. Abreu et al., (DELPHI coll.) Z. Phys. C {\bf 50}, 185 (1991)

\bibitem{OPAL}
P.D. Acton et al., (OPAL coll.) Z. Phys. C {\bf 53}, 539 (1992)

\bibitem{Aleks1}
K.N. Alekseev, Opt. Commun. \textbf{116}, 468 (1995) (preprint
quant-ph/9808010).

\bibitem{Aleks2}
K.N. Alekseev, Pe$\mathrm{\check{r}}$ina J., Phys. Rev. E \textbf{57}, 4023 (1998)
(preprint chao-dyn/9804041).

\bibitem{Aleks3}
K.N. Alekseev, D.S. Prijmak, JETP \textbf{113}, 111 (1998).

\bibitem{Licht}
A.J. Lichtenberg, M.A. Lieberman, \textit{Regular and stochastic motion}
(Springer-Verlag, NY., 1983).

\bibitem{Shuster}
H.G. Shuster, \textit{Deterministic chaos. An Introduction} (Physic-Verlag,
Weinheim, 1984).

\bibitem{Zasl}
G.M. Zaslavsky , P.Z. Sagdeev, \textit{Introduction into nonlinear physics}
(Moscow, 1988). (in Russian)

\bibitem{Krylov}
N.S. Krylov,  \textit{Works on substantiation of statistical physics} (Moscow-Leningrad, 1950).  (in Russian).

\bibitem{Zasl_engl}
G.M. Zaslavsky, Phys. Rep. \textbf{80}, 158 (1991).

\bibitem{Robnic}
T. Prosen, M. Robnik, J. Phys. A\textbf{27}, 8059 (1994).

\bibitem{Mand}
G. Mandelbaum, \textit{Proc. of the Int. Conf. "Chaos and Complexity"} (Edition Frontieres, 1995).

\bibitem{Kuzmin}
V.I. Kuvshinov, A.V. Kuzmin, Proceed. of the Int. Seminar NPCS'01 (Minsk, Belarus, 2001) 183.

\bibitem{Sav1}
G. K. Savvidy, Phys. Lett. B \textbf{71}, 133 (1977).

\bibitem{Kittel}
E. A. De Wolf, I. M. Dremin, W. Kittel, Phys. Rep. \textbf{270}, 1
(1996).
\bibitem{Lin}  
W. A. Lin, L. E. Ballentine, Phys. Rev. Lett. \textbf{65}, 2927 (1990).
\bibitem{Sav}
G. K. Savvidy, S.G. Matinyan, N.G. Ter-Ar.-Savvidy, JETP \textbf{80}, 830 (1981).
\bibitem{Toda}
M. Toda, Phys. Lett. A \textbf{48}, 335 (1974).
\bibitem{Salas}
L. Salasnich, preprint nucl-th/9707035.
\end{thebibliography}
\end{document}